\documentclass[twoside,twocolumn,9pt]{article}
\pdfoutput=1
\usepackage{extsizes}
\usepackage[super,sort&compress,comma]{natbib} 
\usepackage[version=3]{mhchem}
\usepackage[left=1.5cm, right=1.5cm, top=1.785cm, bottom=2.0cm]{geometry}
\usepackage{balance}
\usepackage{times,mathptmx}
\usepackage{sectsty}
\usepackage{graphicx} 
\usepackage{lastpage}
\usepackage[format=plain,justification=justified,singlelinecheck=false,font={stretch=1.125,small,sf},labelfont=bf,labelsep=space]{caption}
\usepackage{float}
\usepackage{fancyhdr}
\usepackage{fnpos}
\usepackage[english]{babel}
\usepackage{array}
\usepackage{droidsans}
\usepackage{charter}
\usepackage[T1]{fontenc}
\usepackage[usenames,dvipsnames]{xcolor}
\usepackage{setspace}
\usepackage[compact]{titlesec}
\usepackage{subcaption}

\usepackage{accents}
\newcommand*{\bigdot}[1]{\accentset{\mbox{\large\bfseries .}}{#1}}

\usepackage{epstopdf}

\definecolor{cream}{RGB}{222,217,201}

\begin{document}

\pagestyle{fancy}
\thispagestyle{plain}
\fancypagestyle{plain}{

\fancyhead[C]{\includegraphics[width=18.5cm]{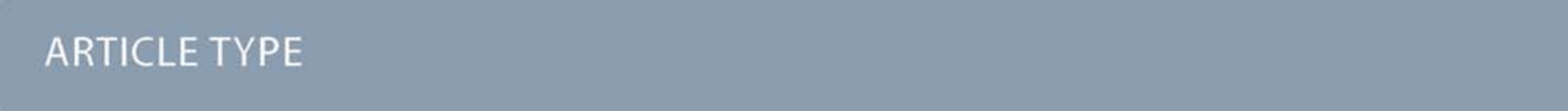}}
\fancyhead[L]{\hspace{0cm}\vspace{1.5cm}\includegraphics[height=30pt]{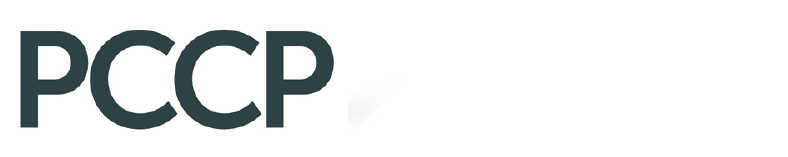}}
\fancyhead[R]{\hspace{0cm}\vspace{1.7cm}\includegraphics[height=55pt]{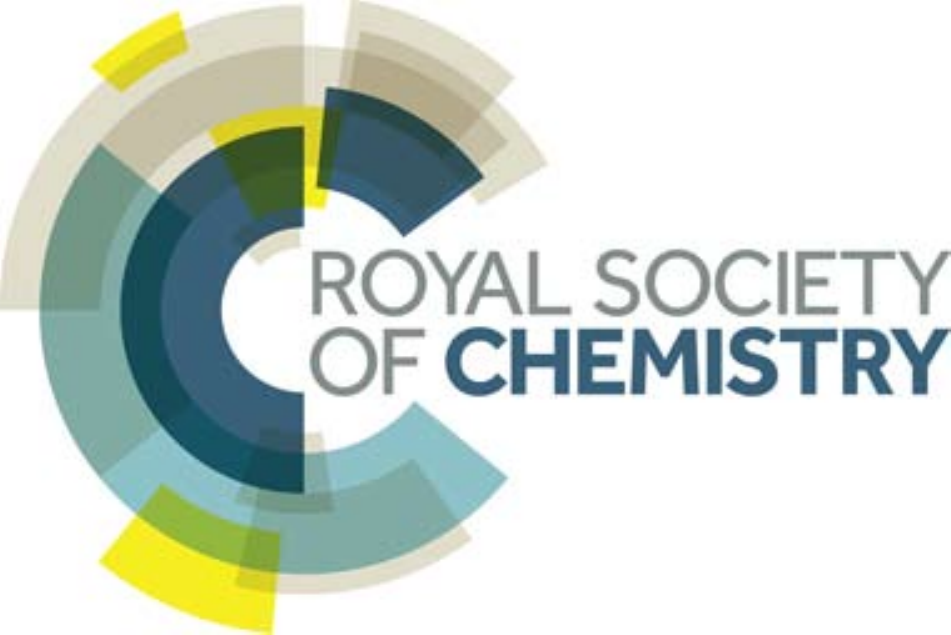}}
\renewcommand{\headrulewidth}{0pt}
}

\makeFNbottom
\makeatletter
\renewcommand\LARGE{\@setfontsize\LARGE{15pt}{17}}
\renewcommand\Large{\@setfontsize\Large{12pt}{14}}
\renewcommand\large{\@setfontsize\large{10pt}{12}}
\renewcommand\footnotesize{\@setfontsize\footnotesize{7pt}{10}}
\makeatother

\renewcommand{\thefootnote}{\fnsymbol{footnote}}
\renewcommand\footnoterule{\vspace*{1pt}%
\color{cream}\hrule width 3.5in height 0.4pt \color{black}\vspace*{5pt}} 
\setcounter{secnumdepth}{5}

\makeatletter 
\renewcommand\@biblabel[1]{#1}            
\renewcommand\@makefntext[1]%
{\noindent\makebox[0pt][r]{\@thefnmark\,}#1}
\makeatother 
\renewcommand{\figurename}{\small{Fig.}~}
\sectionfont{\sffamily\Large}
\subsectionfont{\normalsize}
\subsubsectionfont{\bf}
\setstretch{1.125} 
\setlength{\skip\footins}{0.8cm}
\setlength{\footnotesep}{0.25cm}
\setlength{\jot}{10pt}
\titlespacing*{\section}{0pt}{4pt}{4pt}
\titlespacing*{\subsection}{0pt}{15pt}{1pt}

\fancyfoot{}
\fancyfoot[LO,RE]{\vspace{-7.1pt}\includegraphics[height=9pt]{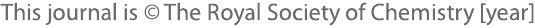}}
\fancyfoot[CO]{\vspace{-7.1pt}\hspace{11.9cm}\includegraphics{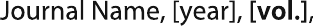}}
\fancyfoot[CE]{\vspace{-7.2pt}\hspace{-13.2cm}\includegraphics{RF}}
\fancyfoot[RO]{\footnotesize{\sffamily{1--\pageref{LastPage} ~\textbar  \hspace{2pt}\thepage}}}
\fancyfoot[LE]{\footnotesize{\sffamily{\thepage~\textbar\hspace{4.65cm} 1--\pageref{LastPage}}}}
\fancyhead{}
\renewcommand{\headrulewidth}{0pt} 
\renewcommand{\footrulewidth}{0pt}
\setlength{\arrayrulewidth}{1pt}
\setlength{\columnsep}{6.5mm}
\setlength\bibsep{1pt}

\makeatletter 
\newlength{\figrulesep} 
\setlength{\figrulesep}{0.5\textfloatsep} 

\newcommand{\topfigrule}{\vspace*{-1pt}%
\noindent{\color{cream}\rule[-\figrulesep]{\columnwidth}{1.5pt}} }

\newcommand{\botfigrule}{\vspace*{-2pt}%
\noindent{\color{cream}\rule[\figrulesep]{\columnwidth}{1.5pt}} }

\newcommand{\dblfigrule}{\vspace*{-1pt}%
\noindent{\color{cream}\rule[-\figrulesep]{\textwidth}{1.5pt}} }

\makeatother

\twocolumn[
  \begin{@twocolumnfalse}
\vspace{3cm}
\sffamily
\begin{tabular}{m{4.5cm} p{13.5cm} }

\includegraphics{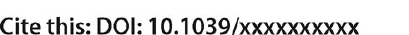} & \noindent\LARGE{\textbf{The influence of absorbing boundary conditions on the transition path times statistics}} \\
\vspace{0.3cm} & \vspace{0.3cm} \\

 & \noindent\large{Michele Caraglio$^{\ast}$\textit{$^{a}$}, 
Stefanie Put\textit{$^{b}$}, 
Enrico Carlon\textit{$^{a}$}, 
Carlo Vanderzande\textit{$^{a,b\ddag}$}} \\

\includegraphics{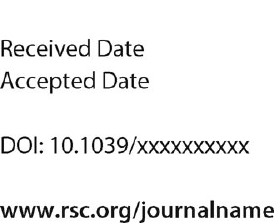} &

\noindent\normalsize{ 
We derive an analytical expression for the transition path time (TPT)
distribution for a one-dimensional particle crossing a parabolic barrier.
The solution is expressed in terms of the eigenfunctions and eigenvalues
of the associated Fokker-Planck equation. The particle performs an
anomalous dynamics generated by a power-law memory kernel, which includes
memoryless Markovian dynamics as a limiting case. Our result takes
into account absorbing boundary conditions, extending existing results
obtained for free boundaries. We show that TPT distributions obtained
from numerical simulations are in excellent agreement with analytical
results, while the typically employed free boundary conditions lead to
a systematic overestimation of the barrier height. These findings may
be useful in the analysis of experimental results on transition path
times. A web tool to perform this analysis is freely available.}\\


\end{tabular}

 \end{@twocolumnfalse} \vspace{0.6cm}

  ]

\renewcommand*\rmdefault{bch}\normalfont\upshape
\rmfamily
\section*{}
\vspace{-1cm}


\footnotetext{\textit{$^{a}$~KU Leuven, Institute for Theoretical Physics, 
Celestijnenlaan 200D, B-3001 Leuven, Belgium.}}
\footnotetext{\textit{$^{b}$~Faculty of Sciences, Hasselt University,
3590 Diepenbeek, Belgium.}}

\footnotetext{\ddag~Corresponding author: carlo.vanderzande@uhasselt.be}




\section{Introduction}

\begin{figure}
\begin{center}
\setlength{\unitlength}{0.5cm}
\begin{picture}(12,5.5)(0,0)
\put(0,0){\makebox(12,5.5){}}

\multiput(3,1)(0,0.4){10}{\line(0,1){0.2}}
\multiput(5.5,1)(0,0.4){10}{\line(0,1){0.2}}

\multiput(2.5,3)(0.4,0){10}{\line(1,0){0.2}}
\multiput(2.5,4)(0.4,0){10}{\line(1,0){0.2}}

\put(6.4,3){\vector(0,1){1}}
\put(6.4,4){\vector(0,-1){1}}

\thicklines
\put(0.2,0.7){\vector(1,0){11.8}}
\put(0.7,0.2){\vector(0,1){5.3}}

\put(2.8,0){$x_a$}
\put(3,0.7){\line(0,1){0.2}}
\put(3,0.7){\line(0,-1){0.2}}
\put(5.3,0){$x_b$}
\put(5.5,0.7){\line(0,1){0.2}}
\put(5.5,0.7){\line(0,-1){0.2}}

\put(11.7,0){$x$}
\put(-1,5){$U(x)$}
\put(6.6,3.3){$\Delta U$}

\put(3,3){\circle*{0.2}}
\put(5.5,3){\circle*{0.2}}

\linethickness{0.4mm}
\qbezier(1.1,5)(2,-2)(3,3)
{\color{red} \qbezier(1.1,5)(2,-2)(3,3)}
\qbezier(3,3)(3.5,5)(5.5,3)
{\color{red} \qbezier(3,3)(3.5,5)(5.5,3)}
\qbezier(5.5,3)(6,2.4)(6.3,1.8)
{\color{red} \qbezier(5.5,3)(6,2.4)(6.3,1.8)}
\qbezier(6.3,1.8)(7.8,-0.6)(10.5,4.3)
{\color{red} \qbezier(6.3,1.8)(7.8,-0.6)(10.5,4.3)}
\end{picture}
\end{center}
\caption{Sketch of a typical energy landscape as a function of a
reaction coordinate $x$. The system fluctuates between the two minima.
Transition paths are the parts of the trajectory originating in
$x_a$ ($x_b$) and ending in $x_b$ ($x_a$) without crossing the two
boundaries.}\label{fig:fig1}
\end{figure}
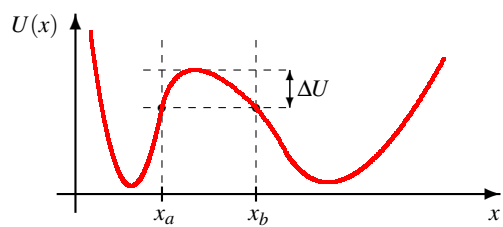

Quite some attention has been devoted in the past years to the study
of transition path times (TPT)~\cite{bere05,dudk06,zhan07,sega07,chun09,
chau10,orla11,neup12,kim15,maka15,true15,dald16,poll16,neup16,sati17,bere17,
lale17,neup17,jana18,carl18,neup18}. In a barrier-crossing process, such
as the folding of biomolecules~\cite{humm04}, transition paths are parts
of a stochastic trajectory corresponding to an actual crossing event.
Biomolecular folding is usually described as a transition between two
stable conformations (the folded and unfolded states) using stochastic
dynamics of a reaction coordinate moving on a double well potential
landscape (see Fig.~\ref{fig:fig1}). Typically, the system spends
most of its time close to one of the two minima, while transitions
have very short duration. Despite the technical challenges due to the
time resolution needed for TPT measurement, several experiments of the
past few years determined average TPT in nucleic acids and protein
folding~\cite{chun09,neup12,true15,neup17,neup18}. The full probability
distribution function of TPT was also obtained~\cite{neup16}.

Transition paths are defined as those trajectories originating at a
point $x_a$ ($x_b$), say, at one side of the barrier and ending in
$x_b$ ($x_a$) at the opposite side (Fig.~\ref{fig:fig1}), without
recrossing $x_a$ ($x_b$). Technically, this corresponds to imposing
absorbing boundary conditions $P(x_a,t)=P(x_b,t)=0$, where $P(x,t)$
is the probability distribution of the particle position at time
$t$. Current stochastic models employed to obtain TPT distributions
use a parabolic barrier~\cite{zhan07,lale17,jana18,carl18}.  As dealing
with absorbing boundaries is challenging, free boundary conditions are
instead preferred~\cite{chun09,lale17}. This is a valid approximation
as long as barriers are high compared to the thermal energy $k_BT$,
since for very high barriers boundaries recrossings are highly unlikely.
However, recent experiments analyzing TPT distributions in nucleic acid
folding~\cite{neup16} estimated a barrier height of $\approx k_BT/2$. This
value was obtained by fitting the data to the analytical form of the TPT
distribution of the free boundary case, but the use of this distribution
for barriers of the order of $k_BT$ or smaller is questionable.

The aim of this paper is to calculate the TPT distribution for
a parabolic barrier imposing absorbing boundaries at the two end
points of the trajectories. This is done for a stochastic system
whose evolution is described by a generalized Langevin equation
with a power-law memory kernel (which also includes the Markovian
dynamics in a limiting case). Anomalous dynamics is ubiquitous
in macromolecular systems as polymers, as it is known from many
examples~\cite{dubb07,saka07,panj07,dubb11,walt12,fred14,vand15,saka17}.
The calculation of the TPT distribution consists in expanding the solution
of the Fokker-Planck equation in an infinite series of eigenfunctions
(confluent hypergeometric functions).  For the numerical estimate
this series is truncated at a sufficiently high order. The results
are in excellent agreement with numerical integration of the Langevin
dynamics. Finally, we developed a web tool performing the fit of TPT
distributions and providing an estimate of barrier height and friction
coefficient in the Markovian case.

\section{Generalized Langevin Equation}

The formalism and theory follow closely the work of Goychuk and
H\"anggi~\cite{goyc07} who considered the case of the escape out of a
cusped-shape parabolic potential.
The starting point is the Generalized Langevin Equation for 
an overdamped particle in a parabolic potential barrier $V(x)=-k x^2/2$
(note that the sign is reversed compared to the escape problem 
discussed in Ref.~\cite{goyc07}):
\begin{equation}
\int_{0}^{t} K(t-\tau) \ \bigdot{x}(\tau) d\tau = k x(t) + \xi(t) \; ,
\label{EQ:GLE}
\end{equation}
where $K(t)$ is a memory kernel and $\xi(t)$ is a random force
with zero average $ \langle \xi(t) \rangle = 0 $ and correlation 
given by the fluctuation-dissipation theorem
\begin{equation}
\langle \xi(t_1)\xi(t_2) \rangle = k_BT K(t_1-t_2) \; ,
\label{EQ:corr_noise}
\end{equation}
In this paper we will consider a power-law memory kernel
\begin{equation}
K(t) = \dfrac{\eta_{\alpha} \, \vert t \vert^{-\alpha}}{\Gamma(1-\alpha)} 
\; , \qquad (0<\alpha<1) \; ;
\label{EQ:Kt}
\end{equation}
with $\eta_{\alpha} = \gamma \, \Gamma(3-\alpha)$.
The Markovian limit is obtained as $\alpha \to 1^-$ leading 
to $K(t_1-t_2)=2 \gamma \delta(t_1-t_2)$, with $\gamma$ the
friction coefficient.

For a parabolic potential \eqref{EQ:GLE} can be mapped onto the following
Fokker-Planck equation~\cite{wang99,adel76,hang77,hyne86,goyc07}
\begin{eqnarray} \label{EQ:FokkerPlanck}
\dfrac{\partial P(x,t)}{\partial t} &=& 
D(t) \dfrac{\partial}{\partial x} \left( e^{\,\beta kx^2/2} 
\dfrac{\partial}{\partial x} e^{-\beta kx^2/2} 
P(x,t) \right) \; ,
\end{eqnarray}
where $\beta = 1/k_BT$ and $P(x,t)$ is the probability density of finding the
particle in position $x$ at time $t$. $D(t)$ is a time-dependent diffusion
coefficient given by~\cite{goyc07}
\begin{equation} \label{EQ:D_t}
D(t) = \dfrac{k_BT}{k} \dfrac{d}{dt} \ln \theta(t) = 
\dfrac{k_BT}{k} \dfrac{\bigdot{\theta}(t)}{\theta(t)} \; ,
\end{equation}
where
\begin{equation} \label{EQ:theta}
 \theta(t) = E_{\alpha} \left[ \dfrac{k}{\eta_{\alpha}} t^{\alpha} \right] \; ,
\end{equation}
with $E_{\alpha}(z) = \sum_{n=0}^{\infty} z^n / \Gamma(\alpha n +1)$
the Mittag-Leffler function. 
In the Markovian limit $\alpha=1$ one has
$E_{\alpha}(z) = e^z$ and, as expected, the time independent Einstein relation,
$D=k_BT/\gamma$, is recovered from~\eqref{EQ:D_t}.

\begin{figure}[t!]
\begin{center}
\includegraphics[scale=1.0]{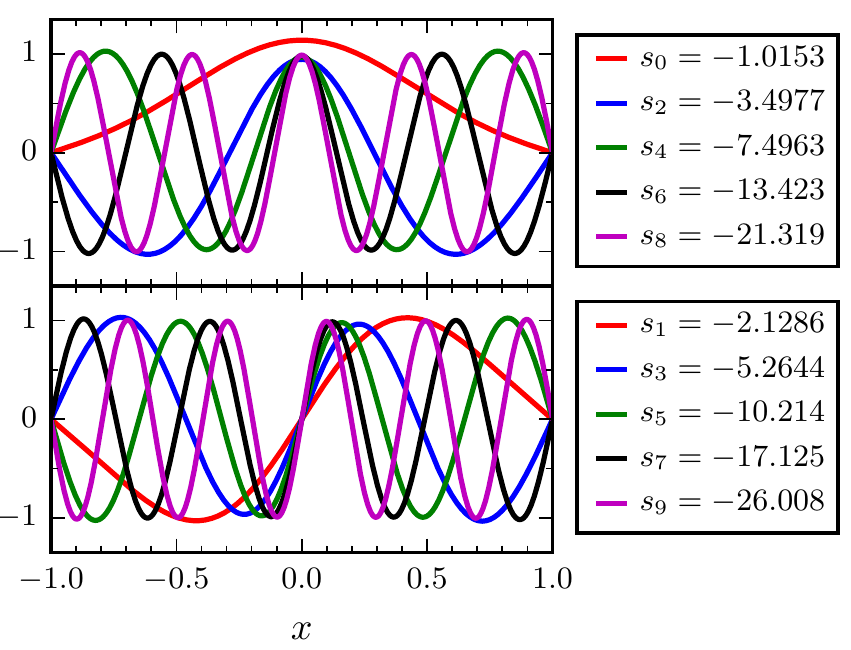}
\caption{Plot of normalized $Y_n(x)$ for $n=0,1,\ldots,9$ with
$\beta=1$ and $k=10$.  Absorbing boundary conditions are imposed in
$x=\pm 1$, which leads to reported values of $s_n$
(obtained from numerical calculations).}
\label{FIG:1F1}
\end{center}
\end{figure}

We seek a solution of Eq.~(\ref{EQ:FokkerPlanck}) in the form
of a spectral expansion
$P(x,t) = \exp(\beta k x^2/4) \sum_n c_n Y_n(x) \varphi_n(t)$.
Separation of variables leads to the following equation for
the position-dependent part
\begin{equation}
Y''_n(x) - \left( \frac{\beta k}{2} + \frac{\beta^2 k^2 x^2}{4} \right) 
Y_n(x)= \lambda_n Y_n(x) \: ,
\label{EQ:Yn}
\end{equation}
while for the time-dependent part
\begin{equation} \label{EQ:coefficientT}
\bigdot{\varphi}_n(t) = \lambda_n D(t) \varphi_n(t) \; .
\end{equation}
The solution of the latter can be easily deduced from~\eqref{EQ:D_t}
\begin{equation} \label{EQ:phi_n}
\varphi_n(t) = [\theta(t)]^{s_n} \; ,
\end{equation}
where $s_n \equiv \frac{k_BT}{k}\lambda_n$. 
Also the solutions of Eq.~\eqref{EQ:Yn} are well known~\cite{abra64}. 
They are either even or odd functions in $x$ and can be written as
\begin{equation} \label{EQ:Solutions_2}
Y_n (x) = 
\left\lbrace
\begin{array}{l}
e^{-\beta kx^2/4} \, {_1}F_1\left( \dfrac{s_n}{2} + \dfrac{1}{2} ; 
\dfrac{1}{2} ; \dfrac{\beta k x^2}{2} \right) 
\qquad \text{n=0,2,4\ldots} \\
\; \\
\sqrt{\beta k}\, x\, e^{-\beta kx^2/4} \, {_1}F_1\left( \dfrac{s_n}{2} + 1 ; 
\dfrac{3}{2} ; \dfrac{\beta k x^2}{2} \right)
\qquad \text{n=1,3,5\ldots}
\end{array}
\right. 
\end{equation}
where $\,_1F_1(a;b;z)$ is the Kummer confluent hypergeometric function.
The boundary conditions $Y_n (\pm x_0) = 0$ fix the allowed values of
$s_n$.  Figure~\ref{FIG:1F1} shows a plot of the first five even and
odd $Y_n(x)$ for $x_0=1$, $\beta=1$ and $k=10$.  The values of $s_n$
for the first $10$ levels are also shown.

We note that Eq.~\eqref{EQ:Yn} is analogous to the Schr\"odinger equation
for a one-dimensional harmonic oscillator. In the ordinary quantum case
one seeks normalizable wavefunctions decaying sufficiently fast at $\pm
\infty$, which leads to the following values $s_n = \lambda_n/\beta k =
-n-1$. One can recognize in this the energy levels of the one-dimensional
quantum oscillator. In the present case, imposing vanishing functions
at some finite $x$, as $Y_n (\pm x_0) = 0$, leads to non-integer values
for $s_n$. Note however (Fig.~\ref{FIG:1F1}) that $s_n$ for the lowest
level are close to the quantum oscillator values $s_n \approx -n-1$.
This is because the lowest states are strongly localized, therefore
imposing vanishing functions at infinity or at some finite $x_0$ leads
to a similar spectrum. The larger is the barrier ($\beta k x_0^2/2$),
the closer is the spectrum of $s_n$ to that of the quantum oscillator.

The most general solution of the Fokker-Planck equation~\eqref{EQ:FokkerPlanck}
can be expressed as a linear combination:
\begin{equation} \label{EQ:spectralexpansion}
P(x,t) = \exp(\beta k x^2/4) \sum_{n=0}^{\infty} c_n Y_n(x) 
\left[\theta(t)\right]^{s_n} \; ,
\end{equation}
where the coefficients $\{c_n\}$ are fixed by the initial conditions.
At $t=0$ particles are placed in $x=-x_0+\varepsilon$, which
corresponds to $P(x,t=0)=\delta(x+x_0-\varepsilon)$ or
\begin{equation} \label{eq_IC}
\delta(x+x_0-\varepsilon) = \exp(\beta k x^2/4)
\sum_{n=0}^{\infty} c_n Y_n(x) \; ,
\end{equation}
since $\theta(0)=1$. The functions $Y_n(x)$ are orthogonal
and we normalize them as follows:
\begin{equation}\label{eq:normalization}
\int_{-x_0}^{+x_0} Y_n^2(x)  dx = 1 \; .
\end{equation}
Multiplying both sides of Eq.~(\ref{eq_IC}) by $Y_m(x)$
and integrating in $-x_0 \leq x \leq x_0$, we get
\begin{equation} \label{EQ:coefficientsC}
c_m = \exp\left[-\frac{\beta k}{4} (x_0-\varepsilon)^2 \right] 
Y_m(- x_0 + \varepsilon) \; ,
\end{equation}
so that Eq.~(\ref{EQ:spectralexpansion}) becomes:
\begin{equation} \label{EQ:spectralexpansion_explicit_2}
P(x,t) = 
\exp\left\{\frac{\beta k}{4}\left[ x^2-(-x_0+\varepsilon)^2\right]
\right\} \sum_{n=0}^{\infty} Y_n(-x_0+\varepsilon)
Y_n(x) [\theta(t)]^{s_n}  \; .
\end{equation}
We note that the coordinate-dependent functions $Y_n(x)$ do not depend on
the friction coefficient $\gamma$ and on the anomalous exponent $\alpha$,
while these affect the time-dependent part $\theta(t)$.

\section{TPT distribution}

The particle current associated to the Fokker-Planck equation is
\begin{equation} \label{EQ:current}
j(x,t) \equiv -D(t) \left( e^{\,\beta kx^2/2} 
\dfrac{\partial}{\partial x} e^{-\beta kx^2/2} 
 P(x,t) \right) \; .
\end{equation}
If we indicate with $j_\varepsilon(x,t)$ the current obtained from
the initial condition $P(x,t=0)=\delta(x+x_0-\varepsilon)$,
the TPT distribution is given by the following relation~\cite{humm04}:
\begin{equation}
P_{TPT}(t)= \lim_{\varepsilon \rightarrow 0} 
\dfrac{\displaystyle j_{\varepsilon}(x_0,t)}
{\displaystyle \int_0^{\infty}j_{\varepsilon}(x_0,t') dt'}  \; .
\label{DEF:TPT}
\end{equation}
Inserting Eq.~(\ref{EQ:spectralexpansion_explicit_2}) into
Eq.~(\ref{EQ:current}) we get
\begin{equation} \label{EQ:current_2}
j_{\varepsilon}(x_0,t) = -D(t) e^{\frac{\beta k}{4}
\left[ x_0^2-(-x_0+\varepsilon)^2\right] } 
\sum_{n=0}^{\infty} Y_n(-x_0+\varepsilon) 
Y'_n(x_0) [\theta(t)]^{s_n} \; ,
\end{equation}
in the limit of small $\varepsilon$ and recalling that the 
boundary condition imposes $Y_n(-x_0)=0$ we have
$Y_n(-x_0+\varepsilon) \approx \varepsilon Y'_n(-x_0)$ and
\begin{equation} \label{EQ:current_3}
j_{\varepsilon}(x_0,t) = -\varepsilon D(t) 
e^{\frac{\beta k}{4}\left[ x_0^2-(x_0-\varepsilon)^2\right] } 
\sum_{n=0}^{\infty} Y'_n(- x_0) Y'_n(x_0) [\theta(t)]^{s_n} \; .
\end{equation}
The current is of order $\varepsilon$, however this factor 
cancels with the normalization constant in~\eqref{DEF:TPT}
hence the probability distribution function in the
limit $\varepsilon \to 0$ remains finite:
\begin{equation}
P_{TPT}(t)=  \dfrac{\displaystyle \sum_{n=0}^{\infty} Y'_n(- x_0) Y'_n(x_0) 
[\theta(t)]^{s_n} D(t)}
{\displaystyle \sum_{n=0}^{\infty} Y'_n(-x_0) Y'_n(x_0)      
\int_0^{\infty} D(t') [\theta(t')]^{s_n} dt'} \; .
\end{equation}
Using Eq.~(\ref{EQ:D_t}) we get
\begin{equation}
 \int_0^{\infty} D(t') [\theta(t')]^{s_n} dt' = \dfrac{-1}{\beta k s_n} \; ,
\end{equation}
which finally yields
\begin{equation} \label{eq_TPTdist}
P_{TPT}(t)= -\dfrac{\displaystyle \sum_{n=0}^{\infty} Y'_n(-x_0) Y'_n(x_0) 
[\theta(t)]^{s_n-1} \bigdot{\theta}(t)}
{\displaystyle \sum_{n=0}^{\infty} Y'_n(-x_0) Y'_n(x_0) \frac{1}{s_n}} \; .
\end{equation}
This probability distribution can be evaluated numerically
once the value of $x_0$ and of the parameters $k$, $\gamma$ and $\alpha$ are known. 
To do that it is convenient to recast Eq.~(\ref{EQ:Solutions_2}) as
\begin{equation} \label{EQ:Solutions_2recasted}
Y_n (x) = 
\left\lbrace
\begin{array}{l}
e^{-\frac{\Delta U}{2} \left( x/x_0 \right)^2} \, 
{_1}F_1\left( \dfrac{s_n}{2} + \dfrac{1}{2} ; \dfrac{1}{2} ; 
\Delta U \left( \dfrac{x}{x_0} \right)^2 \right) 
\qquad \text{n=0,2,4\ldots} \\
\; \\
\sqrt{2 \Delta U}\, \dfrac{x}{x_0}  \, 
e^{-\frac{\Delta U}{2} \left( x/x_0 \right)^2 } \, 
{_1}F_1\left( \dfrac{s_n}{2} + 1 ; \dfrac{3}{2} ; \Delta U \left( \dfrac{x}{x_0} \right)^2 \right)
\: \text{n=1,3,5\ldots}
\end{array}
\right. 
\end{equation}
where $\Delta U = \beta kx_0^2/2$ is the dimensionless barrier height.
Thus the functions $Y_n(x)$ depend only on $\Delta U$ and on the rescaled
coordinate $x/x_0$.

\begin{figure}[t!]
\begin{center}
\includegraphics[scale=1.0]{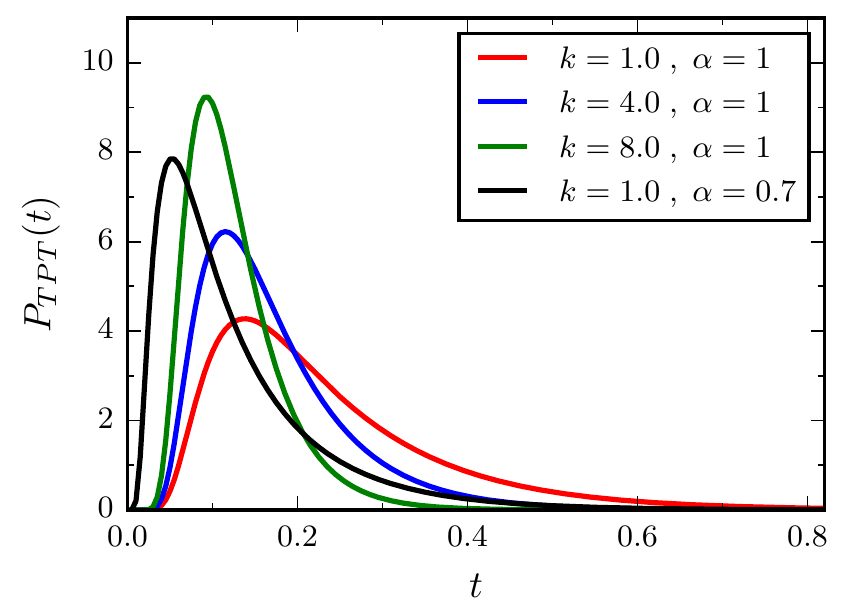}
\caption{Examples of TPT distributions $P_{TPT}(t)$ for $x_0=1$ and
$\gamma = 0.4$. Average TPT decreases with increasing energy barrier and
decreasing $\alpha$. Typically, the distributions are obtained truncating
the infinite series \eqref{eq_TPTdist} to approx $n \leq 100$.}
\label{fig:TPT_exp}
\end{center}
\end{figure}

In the numerical evaluation of Eq.~(\ref{eq_TPTdist}), we chose to
truncate both series in $n$ when the relative increment obtained by
summing one more term is lower than $10^{-6}$.  The values of $s_n$ are
obtained performing the classic Brent method to find a root~\cite{Brent73}
with the initial guess
\begin{equation}
s_n \simeq - \dfrac{\pi^2}{2 \Delta U} \left( \dfrac{n+1}{2} \right)^2 - \dfrac{\Delta U+3}{6} + \mathcal{O} \left( n^{-1} \right) \qquad n=0,1,2,\ldots \; .
\label{eq:sn_asym}
\end{equation}
obtained from the asymptotic properties of hypergeometric
functions~\cite{zeros,NIST:DLMF}.  Figure~\ref{fig:TPT_exp} shows some
plots of Eq.~\eqref{eq_TPTdist} for different $k$ and $\alpha$ and fixed
$x_0=1$ and $\gamma=0.4$.

\begin{figure}[t!]
\begin{center}
\includegraphics[scale=1.0]{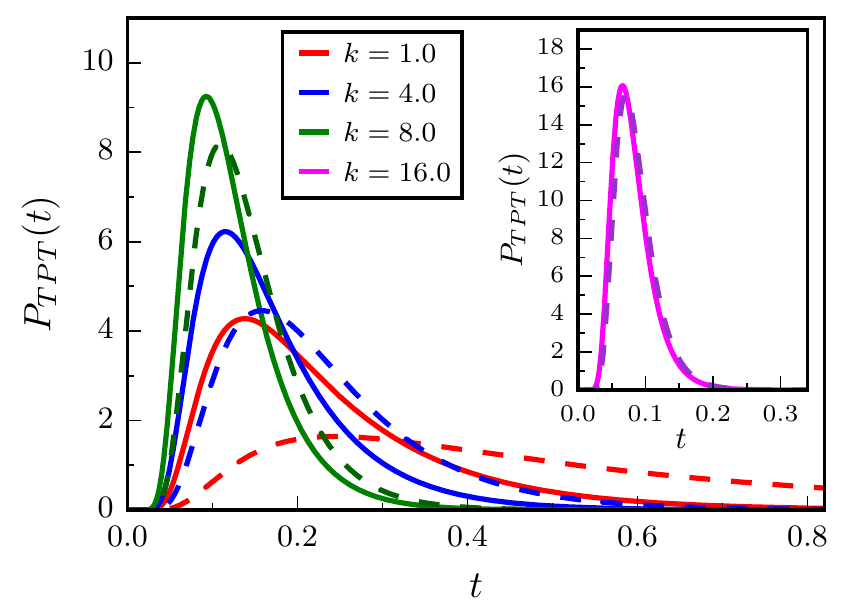}
\caption{Comparison between TPT distributions obtained with absorbing
boundary conditions (Eq.~(\ref{eq_TPTdist}) - solid lines) and free
boundary conditions (Eq.~(\ref{eq_TPTdistNOAB}) - dashed lines).  The two
expressions converge as $k$ increases, corresponding to an increase of
the barrier height. The other parameters are $x_0=1$, $\gamma = 0.4$
and $\alpha=1$.}
\label{fig:TPT_comparison} 
\end{center} 
\end{figure}

\section{The high barrier limit}

The high barrier limit corresponds to the range of rescaled barrier
heights $\Delta U \gg 1$. As mentioned in the Introduction for very
high barriers one expects that TPT distributions are the same whether
either free or absorbing boundary conditions are used. This is because
stochastic trajectories will tend to cross the boundaries at $\pm x_0$
typically only once. The TPT distribution for a parabolic barrier with
free boundaries was shown to be given by~\cite{carl18}
\begin{equation} \label{eq_TPTdistNOAB}
P_{TPT}(t)= -\dfrac{2}{\pi} \dfrac{\bigdot{G}(t) 
e^{-G^2(t)}}{1-Erf[\sqrt{\Delta U}]} \; ,
\end{equation}
where 
\begin{equation} \label{eq_G2}
G^2 (t) \equiv  \Delta U \dfrac{\theta(t)+1}{\theta(t)-1} \; .
\end{equation}
and $\theta(t)$ the Mittag-Leffler function defined in~\eqref{EQ:theta}.
Figure~\ref{fig:TPT_comparison} shows a comparison between
Eq.~\eqref{eq_TPTdistNOAB} (dashed lines) and Eq.~\eqref{eq_TPTdist}
(solid lines). The two expression tend indeed to the same limiting value
as the barrier height increases (increasing $k$).

At long times, the Mittag-Leffler function converges to an
exponential~\cite{haub11} $\theta(t) \sim \exp(\Omega t)$ where we defined $\Omega
\equiv (k/\eta_\alpha)^{1/\alpha}$, which is an intrinsic rate set by
the barrier stiffness $k$ and the noise amplitude $\eta_\alpha$. This
implies that the free boundary distribution \eqref{eq_TPTdistNOAB}
decays asymptotically as~\cite{carl18}
\begin{equation} \label{eq_asymp_free}
P_{TPT}(t) \sim \exp(-\Omega t)
\end{equation}
In the limit $t \to \infty$ the expression~\eqref{eq_TPTdist}
is dominated by the lowest eigenvalue, hence it becomes
\begin{equation} \label{eq_asymp_abc}
P_{TPT}(t) \sim \frac{\bigdot\theta(t)}
{\left[\theta(t)\right]^{1-s_0}} \sim \exp(s_0\Omega t)
\end{equation}
As discussed above, at high barrier $s_n$ converges to the
quantum oscillator energy levels $s_n = -n -1$, which implies
that \eqref{eq_asymp_abc} shows the same asymptotic decay as
\eqref{eq_asymp_free} in the high barrier limit. Note that the exponential
decay \eqref{eq_asymp_abc} remains valid also for low barriers. In this
case the decay rate is $|s_0|\Omega$, i.e.  it differs from the intrinsic
rate $\Omega$.

\section{Fitting empirical distributions}

TPT distributions obtained either from experiments or numerical
simulations can be fitted to Eq.~\eqref{eq_TPTdist}. An online tool
performing the fits is freely available~\cite{online_tool}. In order to
test the procedure we performed some Brownian Dynamics (BD) simulations
for a particle moving in a double-well potential. We restricted the
analysis to the Markovian case $\alpha=1$, which is easier to
handle numerically as the noise is uncorrelated. Note, however, that
the coordinate-dependent functions $Y_n(x)$ and the corresponding values
of $s_n$ (whose calculations are computationally heavy) are independent
of $\alpha$.

\begin{figure}[t!]
\begin{center}
\includegraphics[scale=1.0]{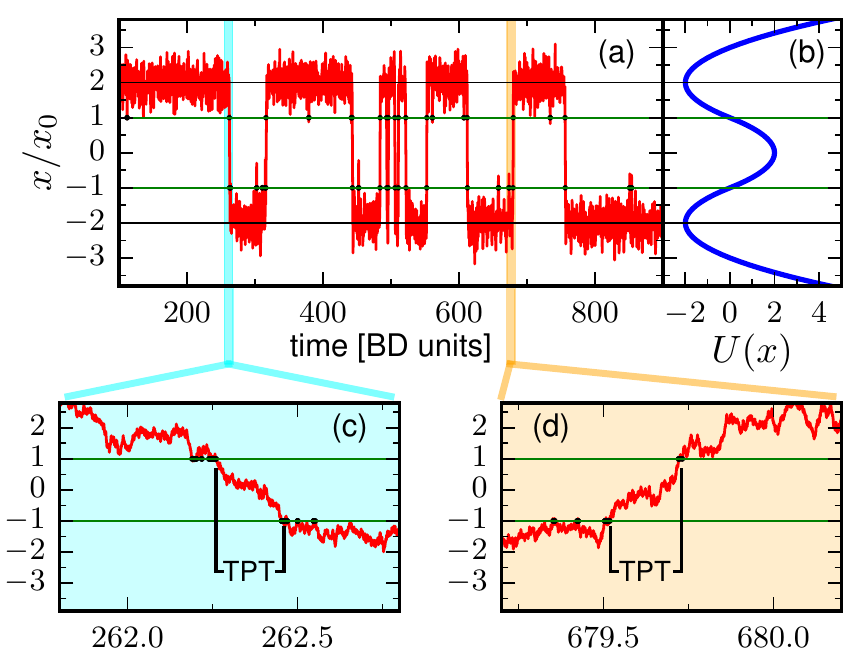}
\caption{a: Example of a BD trajectory of rescaled position $x/x_0$
vs. rescaled time $t/\beta\gamma x_0^2$ (this dimensionless variable
is referred to as BD units in the graph axis). b: Plot of the piecewise
parabolic potential \eqref{EQ:Ux} used in the simulation. c and d:
Details of two parts of the trajectories emphasizing a transition path
from $x_0$ to $-x_0$ (c) and from $-x_0$ to $x_0$ (d).
\label{FIG:BD}}
\end{center}
\end{figure}

We considered the following
piecewise parabolic potential (see Fig.~\ref{FIG:BD}b)
\begin{equation} \label{EQ:Ux}
U (x) = 
\left\lbrace
\begin{array}{l}
k(x+2x_0)^2/2 - kx_0^2/2 \qquad x < -x_0 \\
-kx^2/2 + kx_0^2/2 \qquad -x_0 \leq x \leq x_0 \\
k(x-2x_0)^2/2 - kx_0^2/2 \qquad x > x_0 \\
\end{array}
\right. 
\end{equation}
which has two minima in $x = \pm 2x_0$ and a local maximum in
$x=0$. $U(x)$ and its derivative are continuous, therefore there is no
jump in the force which could be potentially harmful for the simulation.

\begin{figure}[t!]
\begin{center}
\includegraphics[scale=1.0]{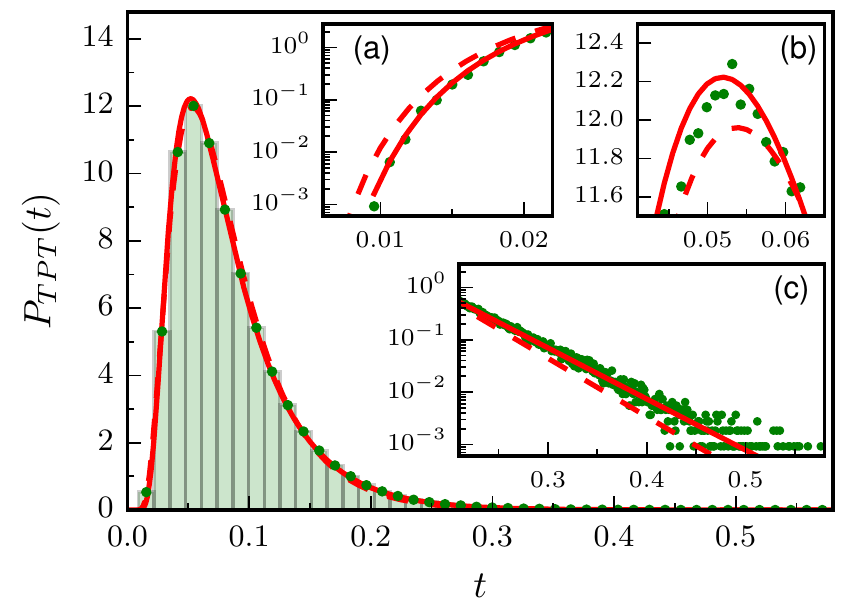}
\caption{Green bars: TPT histogram obtained from BD simulations with
$\Delta U = 1.0$ and $\gamma=0.16$. Red solid line: Best fit of the
simulations to Eq.~\eqref{eq_TPTdist} yielding $\Delta U = 1.0$ and
$\gamma=0.16$ as fitting parameters. Red dashed line: Best fit of the
simulations to Eq.~\eqref{eq_TPTdistNOAB} yielding $\Delta U = 1.57$ and
$\gamma=0.12$ as fitting parameters. Solid and dashed lines are hardly
distinguishable on the scale of the main graph. The insets (a), (b) and
(c) zoom in on the short time behavior, on the maximum and on the long
time behavior of the TPT distribution showing that the solid line fits
better the BD simulations (green circles) compared to the dashed line.
\label{FIG:fitting}}
\end{center}
\end{figure}

Figure~\ref{FIG:BD}a shows an example of a trajectory of rescaled position
$x/x_0$ vs. rescaled time $t/\beta \gamma x_0^2$ (referred to as [BD
units] in the graphs) obtained from BD simulations for a particle moving
in the potential \eqref{EQ:Ux}. The coordinate $x$ fluctuates between
the two minima and the TPT are calculated from the crossings of the
trajectories at $\pm x_0$ (see Fig.~\ref{FIG:BD}c and \ref{FIG:BD}d). We
tested Eq.~\eqref{eq_TPTdist} for several sets of the parameters $k$
and $\gamma$ (with $\alpha=1$).  For each set we collected $10^6$ TPTs,
from which a histogram was obtained. Figure~\ref{FIG:fitting} shows
such a histogram (green bars ending with circles) for $\gamma=0.16$ and
$\Delta U = 1$. The solid red line is obtained by fitting the BD data to
Eq.~\eqref{eq_TPTdist}. The fit provides values of $\gamma$ and $\Delta U$
which are in close agreement to the input values (given in the caption
of Fig.~\ref{FIG:fitting}). The dashed line in Fig.~\ref{FIG:fitting}
is a fit of the BD data to Eq.~\eqref{eq_TPTdistNOAB}. Although the
overall quality of the fit is good a closer look at the short time
scales (inset (a)), the maximum (inset (b)) and the tail (inset (c))
show that Eq.~\eqref{eq_TPTdist} (solid lines) is a better fit to the
data. Moreover the values of $\Delta U$ and $\gamma$ obtained from fitting
the BD simulations to Eq.~\eqref{eq_TPTdistNOAB} deviate sensibly from
the input data. For instance, one gets $\Delta U = 1.57$ which is more
than $50\%$ above the input value.

Table~\ref{tbl:fittingresults} summarizes the results of the fittings
of the BD data. The first two columns give the values of $\Delta U$
and $\gamma$ used in the simulations. The two middle columns are the
outputs of $\Delta U$ and $\gamma$ from fits of Eq.~\eqref{eq_TPTdist}
and the the last two columns give the same parameters from fits of
Eq.~\eqref{eq_TPTdistNOAB}. An uncertainty on the fitted parameters is
also given. The differences in the uncertainties in the two cases is due
to a difference in the fitting procedure. While Eq.~\eqref{eq_TPTdistNOAB}
has a simple analytical form and the search through the parameter space
$(\Delta U,\gamma)$ for optimal fitting parameters is very fast and
efficient, the fitting to Eq.~\eqref{eq_TPTdist} is much more complex.
Each time a pair of values $(\Delta U,\gamma)$ is sampled one needs
to compute the $s_n$ providing the absorbing boundary conditions
$Y_n(\pm x_0)=0$ and this has to be done for a sufficient number of
terms so that one gets an accurate truncation of the infinite series.
To avoid performing these calculations on the fly we considered a grid of
fixed values $\Delta U_k = 0.1 k$ with $1 \leq k \leq 200$ and $\gamma_m
= 0.01 m$ with $1 \leq m\leq 200$.  We note that without any loss of
generality one can set $x_0=1$ since
\begin{equation}
\int_{-x_0}^{+x_0} Y_n^2(x)  dx = x_0 \int_{-1}^{+1} 
\widetilde{Y_n}^2(y) dy \; ,
\end{equation}
where $\widetilde{Y_n} (y) \equiv Y_n (x_0 \, y) $, and that the
integral in the rhs term depends only on $\Delta U$. Hence from the
$x_0=1$ case one can easily generalize the result to arbitrary $x_0$
with a simple rescaling.

\begin{table}[t!]
\small
  \caption{\ 
Fitted values of $\Delta U$ and $\gamma$ on BD simulation data.
The first two columns are the values used in the simulations.  Columns
$3$ and $4$ are obtained from fitting to Eq.~\eqref{eq_TPTdist} while
columns $5$ and $6$ are from fitting to Eq.~\eqref{eq_TPTdistNOAB}.
Typically free boundary conditions (Eq.~\eqref{eq_TPTdistNOAB}) lead to
an overestimate of the barrier height. This does not seem to be true
for the case $\Delta U=4$, which is the highest value of the Table. 
However, it remains true if one fits to
Eq.~\eqref{eq_TPTdistNOAB} while keeping $\gamma$ fixed to the input
value.}
  \label{tbl:fittingresults}
  \begin{tabular*}{0.48\textwidth}{@{\extracolsep{\fill}}cc|cc|cc}
    \hline
    \multicolumn{2}{c|}{BD simulations} & 
    \multicolumn{2}{c|}{Eq.\eqref{eq_TPTdist} - Absorbing BC}  & 
    \multicolumn{2}{c}{Eq.\eqref{eq_TPTdistNOAB} - Free BC} \\
    \hline
    $\Delta U$ & $\gamma$ & $\Delta U \; (\pm 0.1)$ & $\gamma \; (\pm 0.01)$ 
	& $\Delta U \; (\pm 0.01)$ & $\gamma \; (\pm 0.01)$ \\
    \hline
    $0.5$ & $0.16$ & $0.5$ & $0.16$ & $1.30$ & $0.12$ \\
    $0.5$ & $0.40$ & $0.6$ & $0.41$ & $1.29$ & $0.30$ \\
    $0.5$ & $1.20$ & $0.5$ & $1.20$ & $1.29$ & $0.90$ \\
    $1.0$ & $0.16$ & $1.0$ & $0.16$ & $1.57$ & $0.12$ \\
    $1.0$ & $0.40$ & $1.1$ & $0.41$ & $1.56$ & $0.30$ \\
    $1.0$ & $1.20$ & $1.0$ & $1.20$ & $1.56$ & $0.91$ \\
    $2.0$ & $0.16$ & $2.0$ & $0.16$ & $2.16$ & $0.12$ \\
    $2.0$ & $0.40$ & $2.0$ & $0.40$ & $2.16$ & $0.31$ \\
    $2.0$ & $1.20$ & $2.0$ & $1.20$ & $2.16$ & $0.93$ \\
    $4.0$ & $0.16$ & $4.0$ & $0.16$ & $3.61$ & $0.13$ \\
    $4.0$ & $0.40$ & $4.0$ & $0.40$ & $3.57$ & $0.32$ \\
    $4.0$ & $1.20$ & $4.0$ & $1.20$ & $3.58$ & $0.97$ \\
    \hline
  \end{tabular*}
\end{table}

For any sampled pair $(\Delta U_k, \gamma_m)$ the first $1000$ values
of $s_n$ and of $\widetilde{Y_n}'(\pm 1)$ were stored on a database.
Using the stored data one can rapidly estimate the TPT distribution
Eq.~\eqref{eq_TPTdist}. Given an input distribution $\widetilde{P}(t)$
one obtains the optimal fitting values of $\Delta U$ and $\gamma$ by
minimizing over the grid
\begin{equation}
\chi_{k,m} = \int_0^\infty dt               
\left[ \widetilde{P}(t) -
P_{TPT}^{AB}(t; \Delta U_k, \gamma_m) \right]^2 
P_{TPT}^{AB}(t; \Delta U_k, \gamma_m) \: ,
\label{eq:chi}
\end{equation}
where $P_{TPT}^{AB}(t; \Delta U_k, \gamma_m)$ is obtained from
Eq.~\eqref{eq_TPTdist}. Here one minimizes the squared difference of the
two distributions and the multiplication by $P_{TPT}^{AB}(t; \Delta U_k,
\gamma_m)$ ensure that more probable events have higher weight. The values
of the third and fourth columns of Table~\ref{tbl:fittingresults} were
obtained in this procedure, using for $\widetilde{P}(t)$ the distribution
obtained by BD simulations. Obviously, as one uses a fixed grid of
values for the parameters, the accuracy cannot exceed the grid spacing.

In dealing with complex systems, either experimental or numerical,
only a limited sampling is feasible. In order to have some insights on
the error of the parameters and their dependence on the sample size,
we have binned the total $10^6$ TPTs obtained from BD simulations in
subets of sizes $L=10^3$, $L=10^4$ and $L=10^5$. One obtains in this
way $M=10^6/L$ independent samples of size $L$. To estimate the error we
computed the standard deviations $\sigma_{\Delta U}$ and $\sigma_\gamma$
over the values of $\Delta U$ and $\gamma$ for the $M$ samples. The error
was estimated as $\varepsilon_{\Delta U} \equiv \max (\sigma_{\Delta U},
\Delta \Delta U)$ and $\varepsilon_{\gamma} \equiv \max (\sigma_{\gamma},
\Delta \gamma)$, where $\Delta \Delta U=0.1$ and $\Delta \gamma=0.01$ are
the grid spacings used. The data reported in Table~\ref{tbl:fittingerrors}
estimate the typical uncertainty expected for a given sampling size. For
instance using a set of $L=10^4$ TPTs one expects an uncertainty of
$0.1-0.2$ on $\Delta U$, while this error increases for smaller sets.
For a reasonable accuracy on the fitting parameters from the TPT 
distribution one needs about $10^4$ values.

Finally, although this section discussed mainly the fitting of BD
data, the same procedure can be used to fit distributions obtained
from experiments. One just needs to use an experimentally determined
distribution for $\widetilde{P}(t)$ in~\eqref{eq:chi}. An online
server performing these fits is available~\cite{online_tool}.

\begin{table}[t]
\small
  \caption{\ Errors of the parameters evaluated for a given 
number of available TPTs, evaluated as discussed in the main text.
}
  \label{tbl:fittingerrors}
  \begin{tabular*}{0.48\textwidth}{@{\extracolsep{\fill}}cccccccc}
  	\hline
  	 & \multicolumn{3}{c}{$\varepsilon_{\Delta U}$} & 
$\,$ & \multicolumn{3}{c}{$\varepsilon_\gamma$} \\
	$(\Delta U,\gamma)$ & $L=10^5$ \hspace{-8pt} & $L=10^4$  \hspace{-8pt} & $L=10^3$ & & $L=10^5$  \hspace{-8pt} & $L=10^4$  \hspace{-8pt} & $L=10^3$ \\
    \hline
	$(0.5,0.16)$ & 0.1 & 0.14 & 0.30 & & 0.01 & 0.01 & 0.015 \\
	$(0.5,0.40)$ & 0.1 & 0.12 & 0.32 & & 0.01 & 0.01 & 0.027 \\
	$(0.5,1.20)$ & 0.1 & 0.1 & 0.31 & & 0.01 & 0.025 & 0.082 \\
	$(1.0,0.16)$ & 0.1 & 0.14 & 0.44 & & 0.01 & 0.01 & 0.025 \\
	$(1.0,0.40)$ & 0.1 & 0.13 & 0.37 & & 0.01 & 0.01 & 0.029 \\
	$(1.0,1.20)$ & 0.1 & 0.11 & 0.36 & & 0.01 & 0.027 & 0.085 \\
	$(2.0,0.16)$ & 0.1 & 0.15 & 0.45 & & 0.01 & 0.01 & 0.012 \\
	$(2.0,0.40)$ & 0.1 & 0.17 & 0.46 & & 0.01 & 0.011 & 0.031 \\
	$(2.0,1.20)$ & 0.1 & 0.13 & 0.44 & & 0.01 & 0.026 & 0.083 \\
	$(4.0,0.16)$ & 0.1 & 0.2 & 0.6 & & 0.01 & 0.01 & 0.02 \\
	$(4.0,0.40)$ & 0.1 & 0.2 & 0.63 & & 0.01 & 0.013 & 0.03 \\
	$(4.0,1.20)$ & 0.1 & 0.2 & 0.64 & & 0.01 & 0.029 & 0.083 \\
    \hline
  \end{tabular*}
\end{table}

\section{Conclusions}

In this paper we have derived an exact expression for the TPT distribution
for a particle crossing a parabolic barrier in terms of the eigenfunctions
and eigenvalues of the associated Fokker-Planck equation. We considered a
power-law correlated noise characterized by an exponent $\alpha \leq 1$,
see Eqs.~\eqref{EQ:corr_noise} and \eqref{EQ:Kt}, where the Markovian
case is recovered in the limit $\alpha \to 1$. The TPT distribution is
expressed as an infinite series (Eq.~\eqref{eq_TPTdist}) obtained by
separating space and time coordinates, where the coordinate functions
$Y_n(x)$ are independent of the value of $\alpha$. In contrast to
previous work on TPT distributions we have explicitly taken into
account absorbing boundary conditions, which implies $Y_n(\pm x_0)=0$
at the end points of the transition paths $\pm x_0$.  In a recent
work~\cite{sati17} the TPT distribution was determined using a similar
approach in which the eigenfunctions and eigenvalues of a Fokker Planck
equation for a particle in a general potential landscape $V(x)$ were
determined numerically. However in that work it was assumed that $D(t)
\sim t^{\alpha-1}$ a result which is only correct in a linear potential or
for an harmonic one at early times~\cite{footl}.  For other potentials,
the form of $D(t)$ is unknown and one can expect that the form chosen in
\cite{sati17} is valid in an early time regime where the diffusing
particle sees a linear potential. In contrast, we have used the exact
expression for $D(t)$ for a quadratic potential (Eq.~\eqref{EQ:D_t}). 
Note that the coordinate-dependent eigenfunctions do not depend on $D(t)$.

In general, we expect that our result will be very well able to describe
TPT-distributions that come from experiments or simulations on realistic
models of proteins or nucleic acids if at least the transition path times
are measured on intervals where the potential can be well approximated
as a parabola. For this purpose a webtool performing analysis of
TPT distributions has been made freely available~\cite{online_tool}.
Deviations from the result obtained here could be caused by more complex
free energy landscapes where, for example, the reaction coordinate has
to cross more than one barrier or a multidimensional energy landscape.



\balance


\bibliography{bibliography}

\begin{mcitethebibliography}{42}
\bibitem[zer()]{zeros}
For fixed $b$ and $z$, the large $a$-zeros of $\,_1F_1 (a,b,z)$ are given by $$
  a \simeq -\dfrac{\pi^2}{4z} \left[ n^2 + \left(b-\frac{3}{2}\right)n \right]
  + $$ $$ - \dfrac{1}{16z} \left[ \left( b-\frac{3}{2}\right)^2 \pi^2 +
  \frac{4}{3}z^2 - 8b(z-1) - 4b^2 -3 \right] + \mathcal{O}(n^{-1}) \, .$$
  Setting $a=(s_n+1)/2$, $b=1/2$ for $n$ even and $a=s_n/2+1$, $b=3/2$ for $n$
  odd leads to Eq.~\eqref{eq:sn_asym}.\relax
\mciteBstWouldAddEndPunctfalse
\mciteSetBstMidEndSepPunct{\mcitedefaultmidpunct}
{}{\mcitedefaultseppunct}\relax
\EndOfBibitem
\end{mcitethebibliography}
\bibliographystyle{rsc} 

\end{document}